\documentclass[aps,pra,twocolumn,showpacs,superscriptaddress,footinbib,10pt]{revtex4-1}
\usepackage{graphicx}
\usepackage{amsmath}

\newcommand{\be}{\begin{equation}}
\newcommand{\ee}{\end{equation}}

\usepackage{color}
\usepackage[colorlinks,breaklinks,bookmarks=true,citecolor=blue,linkcolor=red,urlcolor=blue]{hyperref}

\begin{document}

\title{Symmetry breaking and unconventional charge ordering in single crystal Na$_{2.7}$Ru$_4$O$_9$}

\author{Arvind Yogi}
\email{yogi.arvind2003@gmail.com}
\affiliation{Center for Correlated Electron Systems, Institute for Basic Science (IBS), Seoul 08826, Korea}
\affiliation{Department of Physics and Astronomy, Seoul National University, Seoul 08826, Korea}

\author{C. I. Sathish}
\affiliation{Center for Correlated Electron Systems, Institute for Basic Science (IBS), Seoul 08826, Korea}
\affiliation{Department of Physics and Astronomy, Seoul National University, Seoul 08826, Korea}

\author{Hasung Sim}
\affiliation{Center for Correlated Electron Systems, Institute for Basic Science (IBS), Seoul 08826, Korea}
\affiliation{Department of Physics and Astronomy, Seoul National University, Seoul 08826, Korea}

\author{Matthew J. Coak}
\affiliation{Center for Correlated Electron Systems, Institute for Basic Science (IBS), Seoul 08826, Korea}
\affiliation{Department of Physics and Astronomy, Seoul National University, Seoul 08826, Korea}

\author{Y. Noda}
\affiliation{J-PARC center, Institute of Materials Structure Science, High Energy Accelerator Research Organization (KEK), Tsukuba, Ibaraki 305-0801, Japan}
\affiliation{Institute of Multidisciplinary Research for Advanced Materials, \linebreak
Tohoku University, Sendai 980-8577, Japan}

\author{Je-Geun Park}
\email{jgpark10@snu.ac.kr}
\affiliation{Center for Correlated Electron Systems, Institute for Basic Science (IBS), Seoul 08826, Korea}
\affiliation{Department of Physics and Astronomy, Seoul National University, Seoul 08826, Korea}

\date{\today }

\begin{abstract}
The interplay of charge, spin, and lattice degrees of freedom in matter leads to various forms of ordered states through phase transitions. An important subclass of these phenomena of complex materials is charge ordering (CO), mainly driven by mixed-valence states. We discovered by combining the results of electrical resistivity ($\rho$), specific heat, susceptibility $\chi$ (\textit{T}), and single crystal x-ray diffraction (SC-XRD) that Na$_{2.7}$Ru$_4$O$_9$ with the monoclinic tunnel type lattice (space group $C$2/$m$) exhibits an unconventional CO at room temperature while retaining metallicity. The temperature-dependent SC-XRD results show successive phase transitions with super-lattice reflections at \textbf{q}$_1$=(0, $\frac{1}{2}$, 0) and \textbf{q}$_2$=(0, $\frac{1}{3}$, $\frac{1}{3}$) below $T_{\textrm{C2}}$ (365 K) and only at \textbf{q}$_1$=(0, $\frac{1}{2}$, 0) between $T_{\textrm{C2}}$ and $T_{\textrm{C1}}$ (630 K). We interpreted these as an evidence for the formation of an unconventional CO. It reveals a strong first-order phase transition in the electrical resistivity at $T_{\textrm{C2}}$ (cooling) = 345 K and $T_{\textrm{C2}}$ (heating) = 365 K. We argue that the origin of the phase transition is due to the localized 4$d$ Ru-electrons. The results of our finding reveal an unique example of Ru$^{3+}$/Ru$^{4+}$ mixed valance heavy \textit{d}$^4$ ions.
\end{abstract}


\maketitle
\section{Introduction}
\label{intro}
Symmetries are important in condensed matter systems with strong ramifications on physical properties. Phase transitions are usually accompanied by a broken symmetry~\cite{Sachdev2011}, which then leads to the appearance of some ordered phases~\cite{Goldstone1962}. In solids, the conduction electrons experience competing interactions with each other and elementary excitations. Ru-based materials have recently emerged as a fertile ground for emergent phenomena driven by a modest spin-orbit coupling (SOC) typical of 4\textit{d} electronic states~\cite{Zhou2016,Carlo2012,Lee2006,Kimber2009,Kimber2012,Lee2013,Grigera2001,Freamat2005}. The 4\textit{d} electrons of Ru are also known to exhibit both localized and itinerant characters. A subtle balance between localization and itinerancy commonly gives rise to a rich variety of electronic and magnetic properties~\cite{Basov2011}.

The heavy $d^4$ ions  based systems have been widely investigated, particularly because of interplay between SOC with intermediate strength ($\simeq$ 0.16 eV) and Coulomb interaction \textit{U}. This competition appears to lead to numerous unusual properties with several interesting examples: unconventional superconductivity~\cite{Miyake1999}, metal-insulator transitions~\cite{Sekine1997,Durairaj2006,Cao2004A}, orbital ordering~\cite{Lee2006}, non-Fermi liquid behaviour~\cite{Cao2004B,Khalifah2001}, high-temperature ferromagnetism~\cite{Longo1968}, low-temperature \textit{p}-wave spin-triplet superconductivity~\cite{Maeno1994}, electron nematic behavior~\cite{Borzi2007,Lester2015}, quantum criticality~\cite{Grigera2001,Rost2009}, and itinerant ferro and metamagnetism~\cite{Mao2006,Cao2003}. Further, the tunnel-type structures such as Rutile, Ramsdellite and Hollandite can accommodate valencies from ${+}$2 to ${+}$5~\cite{Vogt1989,Carter2005} and mixed valencies lead to several interesting phenomena such as charge, spin, and orbital ordering~\cite{Carter1999,Goodenough1955,Rodriguez2005}. In this context, Na$_{2.7}$Ru$_4$O$_9$ (large tunnel type structure) is a promising material for investigating the interplay between Coulomb interactions and the modest SOC of the mixed valance heavy $d^4$ ions.

Among the competing phases, charge ordering is direct evidence of the Coulomb interaction. When the Coulomb interaction is the dominant energy scale, systems tend to be more localized. In this localized state, one often finds an insulating phase with a certain charge ordering (CO). However, this CO becomes unstable as one raises the itinerancy by increasing bandwidth (\textit{W}) or reducing the relative effects of Coulomb \textit{U}. This relative ratio of \textit{W} and \textit{U} gives rise to two routes to the metal-insulator transition (MIT): one is a band-width controlled MIT and the other is Mott-Hubbard MIT. It is generally believed that regardless of the two mechanism of MIT the CO disappears when the system becomes metallic with examples found in 3$d$ transition metal oxides. However, it is not entirely clear how this picture should change with the introduction of spin-orbit coupling. In this sense, it is very important to accumulate enough experimental data before reaching at least some kind of phenomelogical understanding of CO and related MIT for 4$d$ or 5$d$ transition metal oxides (TMO), where the spin-orbit interaction is stronger.

Na$_{2.7}$Ru$_4$O$_9$ crystallizes in a monoclinic crystal structure with a large tunnel running parallel to the crystallographic $b$-axis. The structure consists of corner and edge sharing RuO$_6$ octahedra arranged in one-dimensional zigzag chains, which are formed by single, double and triple edge shared chains, respectively. These zigzag-chains are linked together parallel to the $c$-axis by their corners. In this report, we present physical properties in a wide temperature range (1.9 K $\leq$ $T$ $\leq$ 450 K) using single crystals. We studied the structural, electronic and magnetic properties of the mixed valence Ru$^{3+}$/Ru$^{4+}$ compound Na$_{2.7}$Ru$_4$O$_9$ and report CO behaviour without the loss of metallicity.

\section{Methods}
\label{sec:methods}
Polycrystalline Na$_{2.7}$Ru$_4$O$_9$ samples were synthesized by solid state reaction of preheated RuO$_{2}$ ($99.999$\%, Aldich) and Na$_{2}$CO$_{3}$ ($99.999$\%, Aldich) under an Ar-gas environment at $900$\,$^{\circ}$C for 72 h with several intermediate grindings and pelletizations. Subsequently, high-quality single crystals of Na$_{2.7}$Ru$_4$O$_9$ were grown from this polycrystalline powder via a modified self-flux vapor transport reaction under flowing Ar-gas (ultra-pure $99.999$\%). Long needle shaped high-quality single crystals (1 $\times$ 0.1 $\times$ 0.1 mm$^3$) were obtained from the final products.

The phase purity and temperature-dependent (300 to 450 K) powder XRD were performed by using a Bruker D8 Discover diffractometer with a Cu-K$_\alpha$ source with no impurity peaks observed. An elemental analysis was subsequently done confirming the stoichiometry of the samples: we used a COXI EM-30 scanning electron microscope equipped with a Bruker QUANTAX 70 energy dispersive x-ray system. The temperature-dependent single crystal XRD (SC-XRD) was performed from 300 to 695 K by using a single crystal diffractometer (XtaLAB P200, Rigaku). The crystal structure was refined using both powder and single crystal XRD data with the \texttt{Fullprof}~\cite{fullprof} software suite.

\begin{figure}
\includegraphics[width=0.80\linewidth]{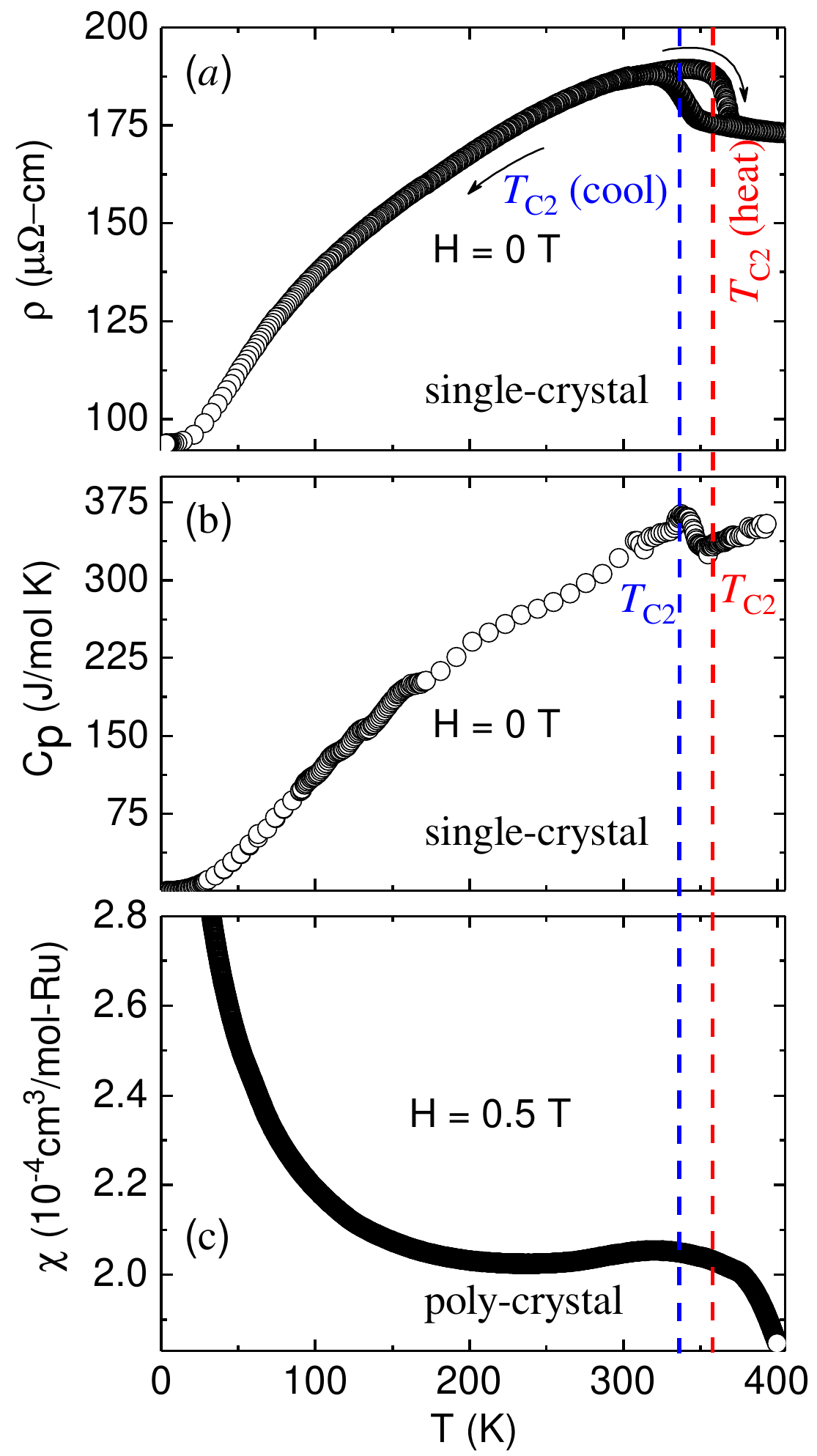}
\caption{\label{fig:FOT}
(Color online). The first order phase transition is observed at $T_{\textrm{C2}}$ (cooling) = 345 K and $T_{\textrm{C2}}$ (heating) = 365 K, with a clear hysteresis in resistivity ($\rho$) upon warming and cooling (a), which is also evident in specific heat (C$_p$) (b) and in magnetic susceptibility $\chi$ (\textit{T}) of Na$_{2.7}$Ru$_4$O$_9$ (c).}
\end{figure}

Electrical resistivity ($\rho$) measurements were carried out using a home-made system equipped with a furnace (300 to 450 K) and a pulsed-tube cryostat (down to 3 K, Oxford). The electrical resistance was measured in the four-point geometry on a static sample holder, where the contacts to the sample were made using silver paint and 25 $\mu$m gold wire. Current ($I$) was applied perpendicular to the single crystal length, which is the crystallographic $b$-axis. Magnetic susceptibility $\chi(T)$ measurements were taken using a MPMS-SQUID magnetometer (Quantum Design). Heat capacity $C_p(T)$ measurements were made using the commercial Physical Property Measurement System (PPMS, Quantum Design).

\begin{figure*}
\includegraphics[width=1.0\linewidth]{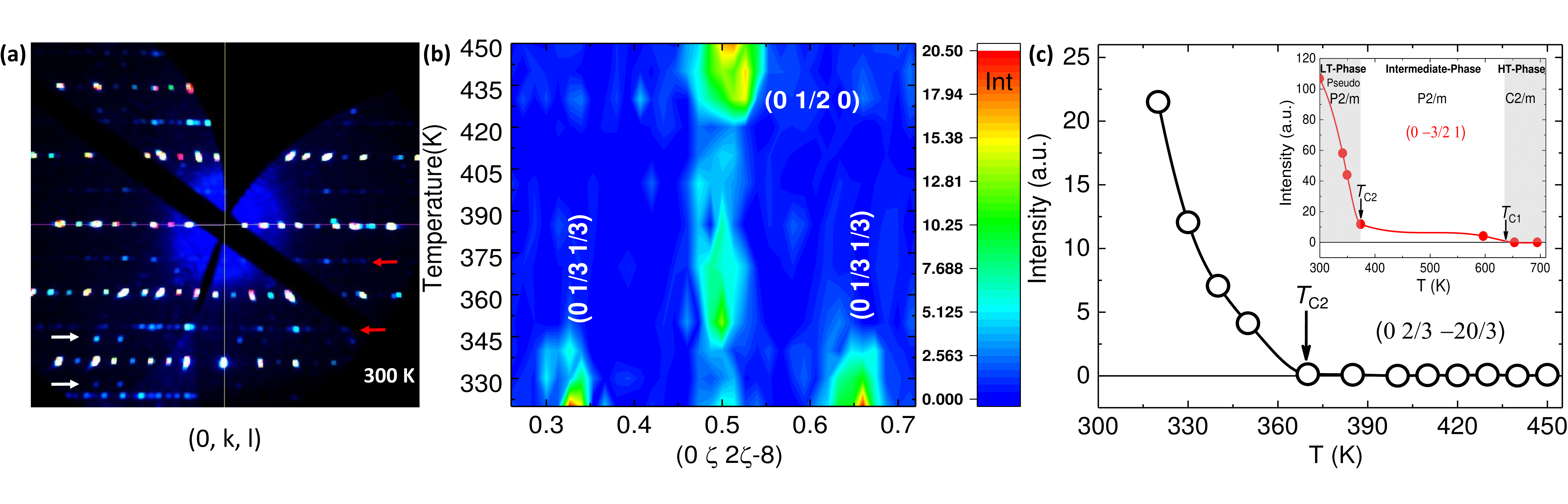}
\caption{\label{fig:line-cut}
(Color online) (a) Reciprocal lattice map of the SC-XRD data for Na$_{2.7}$Ru$_4$O$_9$ measured at 300 K below the first order transition temperature $T_{\textrm{C2}}$. It shows the fundamental ($0, k, l$) reflections (white intense spots) and the satellite weak reflections corresponding to super-lattice peaks at \textbf{q}$_1$=(0, $\frac{1}{2}$, 0) and \textbf{q}$_2$=(0, $\frac{1}{3}$, $\frac{1}{3}$) in the $b^*$-$c^*$ plane, marked by red and white arrows, respectively. (b) The line cut for various temperatures below and above the first order transition as extracted from the reciprocal image analysis exhibits two super-lattice peaks at \textbf{q}$_1$=(0, $\frac{1}{2}$, 0) and \textbf{q}$_2$=(0, $\frac{1}{3}$, $\frac{1}{3}$). (c) The super-lattice peaks at \textbf{q}$_2$=(0, $\frac{1}{3}$, $\frac{1}{3}$) is shown as a function of temperature, which disappears above the first order transition ($>$ $T_{\textrm{C2}}$) $\approx$~370 K. Above $T_{\textrm{C2}}$, only the \textbf{q}$_1$=(0, $\frac{1}{2}$, 0) super-lattice peak was observed, which also disappear above $T_{\textrm{C1}}$ as shown in the inset.}
\end{figure*}

\section{Results}
\label{sec:results}
\subsection{Electrical resistivity, Heat-capacity and Magnetization}
\label{sec:FOT}

Electrical resistivity is shown for single crystal Na$_{2.7}$Ru$_4$O$_9$ in Fig.~\ref{fig:FOT} (a) in the temperature range of 3.3 K~$\leq$~$T$~$\leq$ 450 K for both heating and cooling cycles. Metallic conductivity (d$\rho$/d\textit{T}~$>$~0) is observed in the entire temperature range with a room temperature resistivity of 187.25 $\mu\Omega$-cm. The electrical resistivity reveals a clear hysteresis, indicative of a first-order phase transition with two transitions at $T_{\textrm{C2}}$ (cooling) = 345 K and $T_{\textrm{C2}}$ (heating) = 365 K: which is corroborated well by magnetization and heat-capacity results (Fig.~\ref{fig:FOT}). However, no significant difference could be observed for the resistivity up to 9 T as compared with the zero field resistivity. In order to shed light on the scattering mechanism involved and to have a quantitative understanding of the measured results, the electrical resistivity data of Na$_{2.7}$Ru$_4$O$_9$ was analyzed theoretically by using a Bloch-Gr$\ddot{u}$neisen-Mott model~\cite{Pikul2003}. Using this model, we can show that the temperature dependence of the resistivity can be explained by the electron-phonon and inter-band electron mediated scattering mechanisms [see supplementary information (SI-1) and Fig. S1].

Heat capacity for Na$_{2.7}$Ru$_4$O$_9$ was measured in the temperature range 1.9 K $\leq$ $T$ $\leq$ 400 K and displayed in Fig.~\ref{fig:FOT} (b). The specific heat shows two strong kinks with an inflection point at $T_{\textrm{C2}}$ (cooling) = 345 K and $T_{\textrm{C2}}$ (heating) = 365 K as shown by dashed lines in Fig.~\ref{fig:FOT} (b). The sharp feature of the peak, different from the usual lambda-like shape, confirms a first-order transition in Na$_{2.7}$Ru$_4$O$_9$. The absence of a clear sign of magnetic ordering in the susceptibility shown in Fig.~\ref{fig:FOT} (c) indicates that a structural transition is most likely responsible for the observed hysteresis in the resistivity. The low temperature heat capacity data can be analyzed in terms of
\begin{equation}
 C_P=\gamma T~+~\beta T^3,
\end{equation}
where $\gamma$ is the electronic specific-heat coefficient and $\beta$ is the phonon contribution to the total specific-heat~\cite{Regan2005}. The evidence for a metal-like electronic contribution to the low-temperature heat capacity was observed. The electronic contribution to the specific heat ($\gamma$) for Na$_{2.7}$Ru$_4$O$_9$ was determined to be 26.91 mJ/mol K$^2$, indicating an enhanced contribution of conduction electrons in excellent agreement with the transport measurements. Moreover, the estimated value of $\gamma$ for Na$_{2.7}$Ru$_4$O$_9$ is much larger than the free electron value of $\gamma$ and comparable with the values for other ruthenates: e.g. Ru superconductor Sr$_2$RuO$_4$ ($\gamma$ = 40 mJ/mol K$^2$ Ru) and non-Fermi-liquid compound La$_4$Ru$_6$O$_{19}$ ($\gamma$ = 25 mJ/ mol K$^2$ Ru)~\cite{Maeno1994,Khalifah2001}.  This large value of $\gamma$ indicates that the conduction electrons, most likely from Ru 4$d$ bands, are strongly correlated.

Bulk magnetic susceptibility ($\chi$) for Na$_{2.7}$Ru$_4$O$_9$ was measured as a function of temperature in the temperature range 1.9 K $\leq$ $T$ $\leq$ 400 K under an applied field 0.5 T using a poly-crystalline sample as shown in Fig.~\ref{fig:FOT} (c). Magnetization as a function of applied field ($H$) was measured at 2 and 400 K, exhibits linear variation without any hysteresis with a maximum applied field of 7 T. For practical reasons, we measured magnetization using a large mass of crushed powder sample. The magnetic susceptibility (1.9 K $\leq$ $T$ $\leq$ 400 K) of the compound is temperature independent from room temperature to 150 K, indicative of Pauli-paramagnetic behaviour, which is followed by a Curie-like increase at lower temperature. Notably, the susceptibility curve above room temperature demonstrates a broad hump at the first order phase transition as seen in Fig.~\ref{fig:FOT} (c). This is probably due to short range correlations present at such high temperatures.

We must note that bulk properties of Na$_{2.7}$Ru$_4$O$_9$ are significantly different from Na$_{2}$Ru$_4$O$_{9-\delta}$, which shows large anisotropy in both resistivity and magnetic susceptibility ($\chi$). The metallic as well as magnetic behaviour of Na$_{2}$Ru$_4$O$_{9-\delta}$ was attributed to the presence of localized as well as itinerant Ru electrons~\cite{Cao1996}.

\subsection{X-ray diffraction measurements}
\label{sec:xrd}

\begin{figure}
\includegraphics[width=1.0\linewidth]{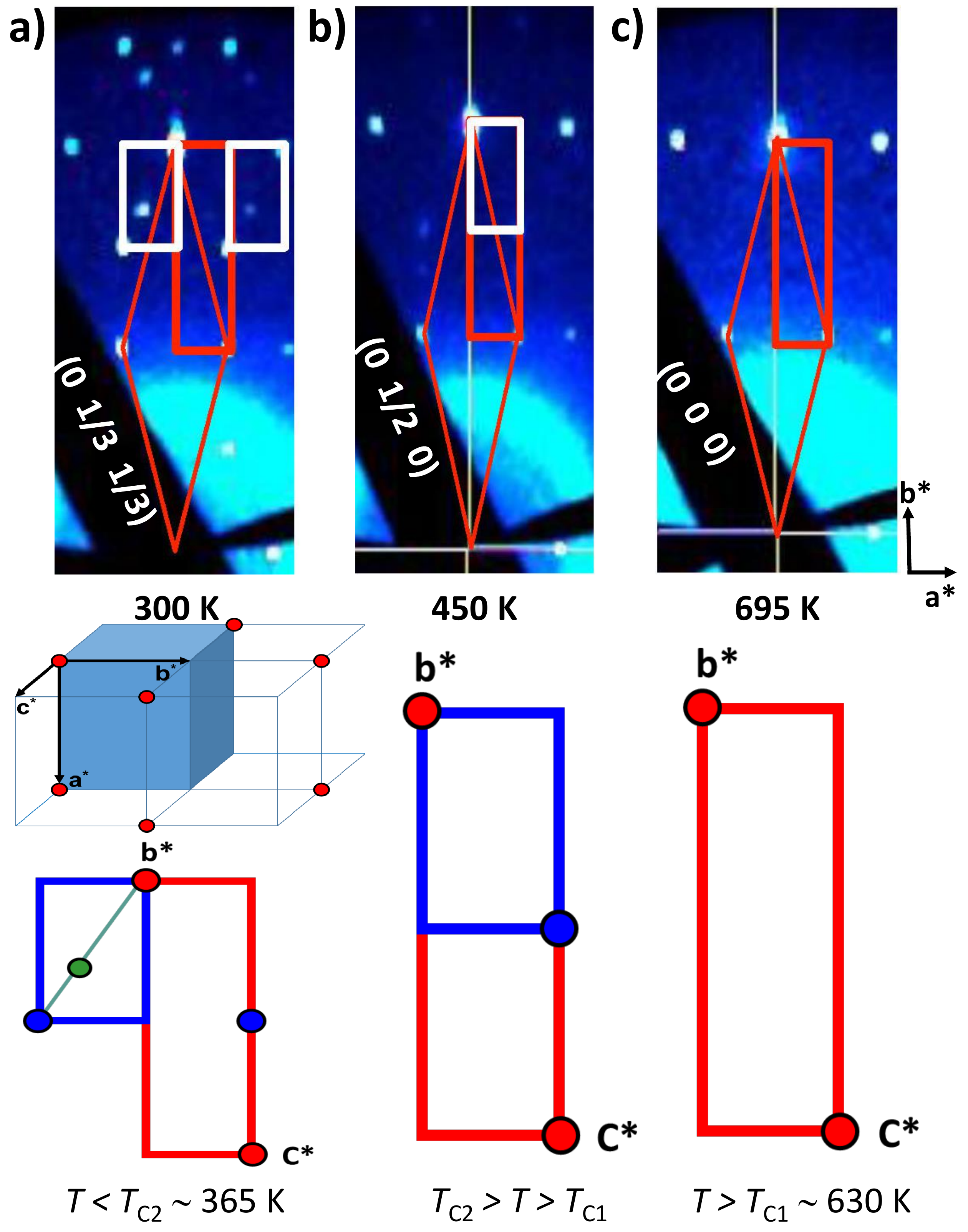}
\caption{\label{fig:structural-model} (Color online) Schematic diagrams of reciprocal unit-cell at the LT-phase at 300 K (a) and at the intermediate-phase at 450 K (b) including super-lattice peaks at \textbf{q}$_1$=(0, $\frac{1}{2}$, 0) and \textbf{q}$_2$=(0, $\frac{1}{3}$, $\frac{1}{3}$) as observed in SC-XRD measurements. The SC-XRD data with schematic diagrams above $T_{\textrm{C1}}$ at the final HT-phase at 695 K (c), where no superstructure reflections were observed (prototype $\equiv$~$C$2/$m$). The unit-cell of the LT-phase (300 K) and the intermediate-phase (450 K) becomes ($a_0$ $\times$ 6$b_0$ $\times$ 3$c_0$) and ($a_0$ $\times$ 2$b_0$ $\times$ $c_0$), respectively as compared to the prototype HT-phase $C$2/$m$ unit-cell ($a_0$ $\times$ $b_0$ $\times$ $c_0$).}
\end{figure}

In order to understand the origin of the phase transition observed in the electrical resistivity and heat capacity data, we performed multiple diffraction experiments on single crystal and poly-crystalline Na$_{2.7}$Ru$_4$O$_9$ in the temperature range between 300 and 695 K. Interestingly enough the temperature-dependent SC-XRD results show successive phase transitions. For example, a single super-lattice peak appears at \textbf{q}$_1$=(0, $\frac{1}{2}$, 0) in the intermediate-phase ($T_{\textrm{C1}}$ $\equiv$ 630 K), whereas two super-lattice peaks emerge at \textbf{q}$_1$=(0, $\frac{1}{2}$, 0) and \textbf{q}$_2$=(0, $\frac{1}{3}$, $\frac{1}{3}$) below $T_{\textrm{C2}}$ in the LT-phase as shown in Fig.~\ref{fig:line-cut} (a and b). We note that these super-lattice peaks were not reported in the previous study~\cite{Regan2005} on powder samples, most probably due to weak reflections.

In addition, the super-lattice peak at \textbf{q}$_2$=(0, $\frac{1}{3}$, $\frac{1}{3}$) disappears on heating as shown in Fig.~\ref{fig:line-cut} (b and c). The line cut from the temperature-dependent reciprocal image analysis for the peak \textbf{q}$_2$=(0, $\frac{1}{3}$, $\frac{1}{3}$) shows a clear suppression of intensity above $T_{\textrm{C2}}$ (Fig.~\ref{fig:line-cut} (c)). This behavior is similar to that observed in Na deficient Na$_{3-x}$Ru$_4$O$_9$~\cite{Onodaa2000}, where Na$^+$-NMR spectra results show motional averaging of the Na$^+$ sites at 390 K (above the first order phase transition). According to Ref.~\cite{Onodaa2000}, this motional averaging of the Na$^+$ sites was discussed in the context of the ionic motion in the material. And above 360 K the ordered state begins to melt rapidly, consistent with our transport results. This suggests a large variation in the charge separation patterns in a mixed valance system below and above the melting of the ordered state (see Section:~Crystal-structure analysis).

The super-lattice spots indicate that the observed modulations in Na$_{2.7}$Ru$_4$O$_9$ are commensurate (\textbf{q}$_1$=(0, $\frac{1}{2}$, 0) and \textbf{q}$_2$=(0, $\frac{1}{3}$, $\frac{1}{3}$)) super-structure type with translation and rotational symmetry breaking. Interestingly, this symmetry breaking induces large ionic displacements (Na ions) in the lattice, which can be explained by a monoclinic structure with the subgroup of $C$2/$m$. This $C$2/$m$ unit-cell doubles the cell ($a_0$ $\times$ 2$b_0$ $\times$ $c_0$) with \textbf{q}$_1$=(0, $\frac{1}{2}$, 0) below 630 K and with further cooling makes the cell 18 times larger ($a_0$ $\times$ 6$b_0$ $\times$ 3$c_0$) with another \textbf{q}$_2$=(0, $\frac{1}{3}$, $\frac{1}{3}$) from the prototype HT-phase. The schematic diagrams of the reciprocal unit-cell are shown in Fig.~\ref{fig:structural-model} at room temperature (\textbf{q}$_1$=(0, $\frac{1}{2}$, 0) and \textbf{q}$_2$=(0, $\frac{1}{3}$, $\frac{1}{3}$)), at 450 K (\textbf{q}$_1$=(0, $\frac{1}{2}$, 0)) and above the $T_{\textrm{C1}}$ at 695 K with reciprocal lattice maps, respectively. The mirror and two-fold symmetry were not observed in the SC-XRD data in the LT-phase (300 K). Therefore we can conclude that the observed Laue symmetry should be $P\overline{1}$.

Further, the super-lattice peak at \textbf{q}$_1$=(0, $\frac{1}{2}$, 0) appears at $T_{\textrm{C1}}$ and suddenly increases again below $T_{\textrm{C2}}$ as shown in the inset of Fig.~\ref{fig:line-cut} (c). The observed behavior of this super-lattice peak reflects the large displacements of Na ions below $T_{\textrm{C2}}$ (see Section:~Displacement pattern). However, it is generally believed that such weak additional satellite reflections are due to the presence of various type of charge density wave (CDW) or CO states in tunnel type structures such as Hollandite A$_x$M$_8$O$_{16}$~\cite{Carter2005}.

\subsection{{Crystal-structure analysis}
\label{sec:x-ray}}

\begin{figure}
\includegraphics[width=1\linewidth]{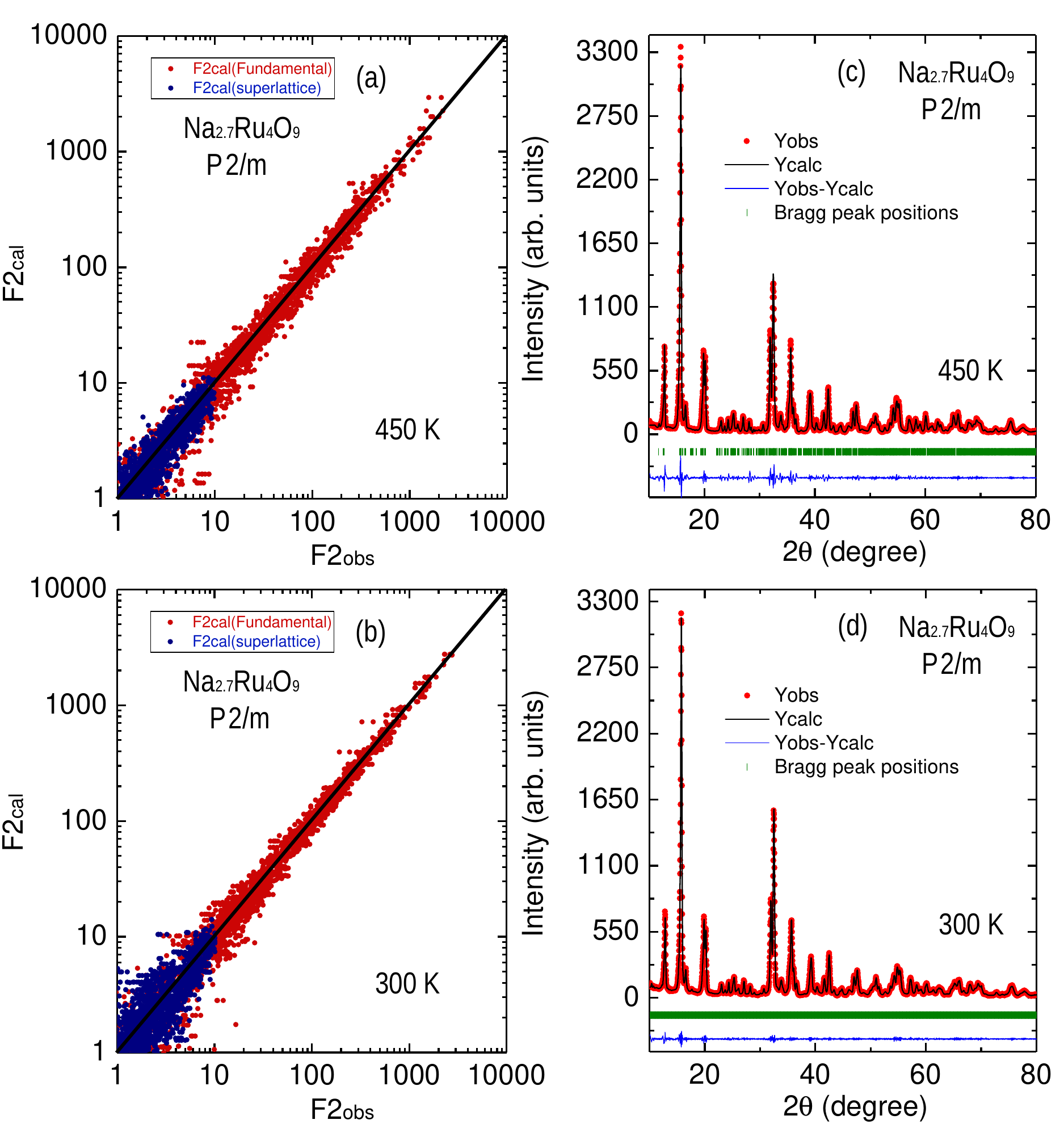}
\caption{\label{fig:XRD-refinement}(Color online). The left panel shows plots for F$^{2}$-obs and F$^{2}$-cal for single crystal experiments at (a) 450 and (b) 300 K, respectively. Red markers represent fundamental reflections, while blue markers represent super-lattice reflections. Note that the vertical and horizontal axis are given in a logarithmic scale. The right panel shows Le-Bail refined x-ray powder diffraction patterns at (c) 450 and (d) 300 K, respectively.}
\end{figure}

The crystal structure of Na$_{2.7}$Ru$_4$O$_9$ has been investigated at intermediate temperature (450 K) and room-temperature (300 K) SC-XRD as shown in Fig.~\ref{fig:XRD-refinement} (a and b) and further confirmed by powder-XRD measurements Fig.~\ref{fig:XRD-refinement} (c and d). We observed a single super-lattice peak with \textbf{q}$_1$=(0, $\frac{1}{2}$, 0) for the intermediate-phase (450 K) and two super-lattice peaks with \textbf{q}$_1$=(0, $\frac{1}{2}$, 0) and \textbf{q}$_2$=(0, $\frac{1}{3}$, $\frac{1}{3}$) for the LT-phase (300 K) in the SC-XRD data (Fig.~\ref{fig:structural-model}). These super-lattice peaks can be indexed by using the monoclinic space group $C$2/$m$ of the prototype HT-phase. Under the group-subgroup relation, $P$2/$m$ is assigned to the intermediate \textbf{q}$_1$=(0, $\frac{1}{2}$, 0) phase. The \textbf{q}$_2$=(0, $\frac{1}{3}$, $\frac{1}{3}$) super-lattice reflection leads to a significant increase of the unit cell volume and lowers the symmetry drastically at room temperature. To explain the \textbf{q}$_1$=(0, $\frac{1}{2}$, 0) super-lattice peak for the intermediate-phase at 450 K, $P$2/$m$ symmetry was used to refine the SC-XRD data under the group-subgroup relation by taking atomic positions from the earlier report on powder samples~\cite{Regan2005} at room room temperature with $C$2/$m$ space group (corresponding to prototype HT-phase, $>$~630 K). The structure of the room temperature phase with \textbf{q}$_1$=(0, $\frac{1}{2}$, 0) and \textbf{q}$_2$=(0, $\frac{1}{3}$, $\frac{1}{3}$) is more complicated. From the observed Laue symmetry, we conclude that the space group should be $P\overline{1}$, but there are too many fitting parameters. Thus, we ignored the \textbf{q}$_2$=(0, $\frac{1}{3}$, $\frac{1}{3}$) super-lattice reflections and made an averaged structure analysis with the same unit-cell size and the same space group $P$2/$m$ of the intermediate-phase as an approximant structure. The obtained R-factors ($R_{F2}$) for the intermediate and LT-phases are 11.41 and 16.01, respectively (see Fig.~\ref{fig:XRD-refinement} (a and b)). The refined lattice parameters for the prototype HT-phase at 695 K, the intermediate-phase at 450 K and the LT-phase at 300 K (room temperature) are shown in Table ~\ref{tab:cell}.

\begin{figure}
\includegraphics[width=0.80\linewidth]{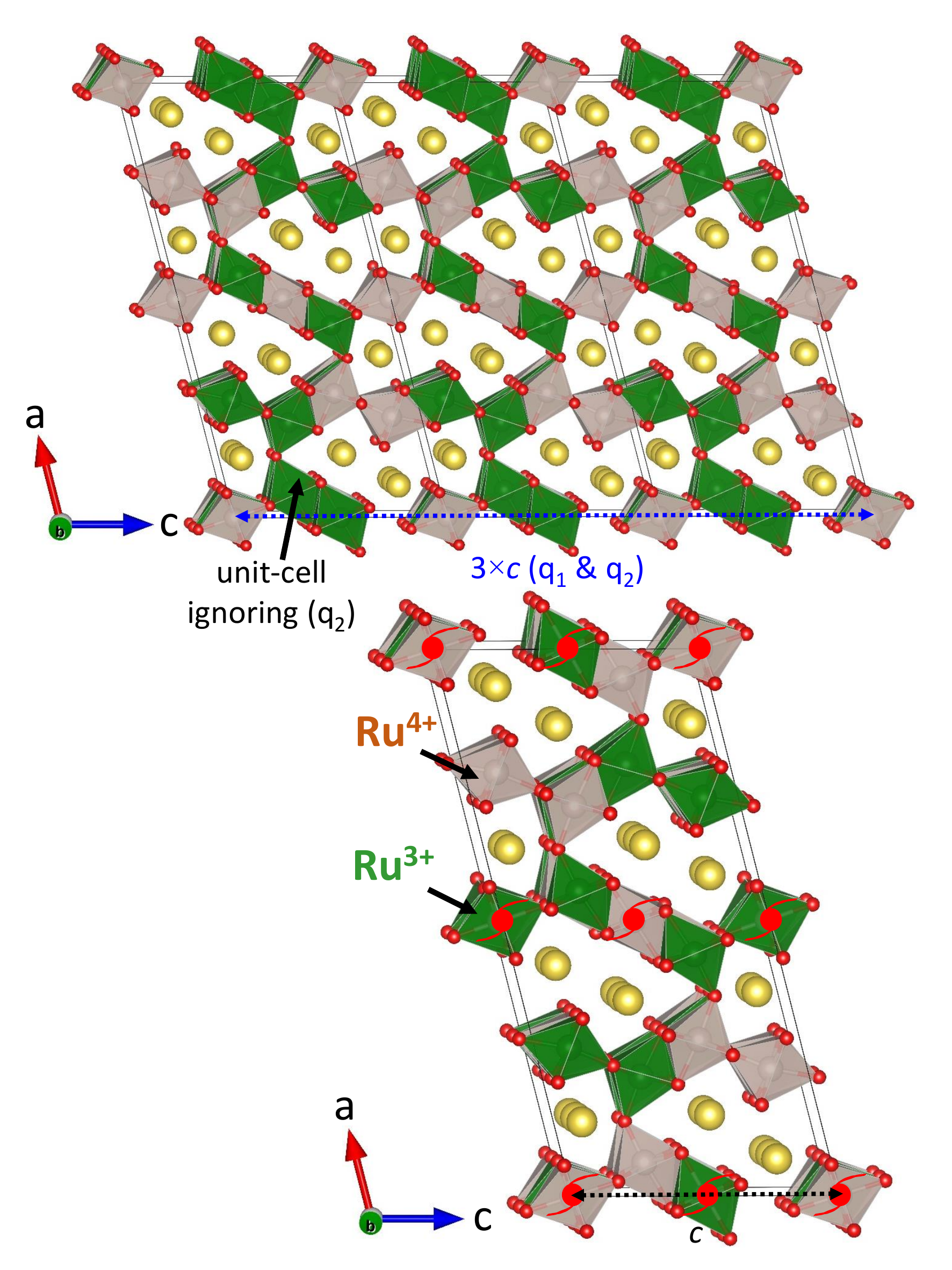}
 \caption{\label{fig:BVS} (Color online) The projection of crystal structure of Na$_{2.7}$Ru$_4$O$_9$ at the intermediate-phase (450 K) (bottom panel) and at the LT-phase (300 K) (top panel) in the crystallographic $ac$-plane. Unit cell at 300 K is simply tripled from the analyzed cell on drawing. BVS calculation at 450 and 300 K shows the segregation of Ru-valence as Ru$^{3+}$ (green) / Ru$^{4+}$ (light brown) RuO$_6$ octahedral.}
\end{figure}

The crystal structure of Na$_{2.7}$Ru$_4$O$_9$ for the intermediate and low (room) temperature phases is shown in Fig.~\ref{fig:BVS}. The structural parameters obtained from SC-XRD refinement are shown in Table S1 and S2 (see SI-2) and were used as the starting point of powder HR-XRD structural refinements to double check our structural model. All Bragg reflections were well indexed by assuming the monoclinic space group $P$2/$m$ and the HR-XRD refinement shows an excellent fit as shown in Fig.~\ref{fig:XRD-refinement} (c and d). Note that super-lattice reflections found by single-crystal experiments are too weak to be seen in the powder reflection profiles shown in Fig.~\ref{fig:XRD-refinement} (c and d).

To explain the crystal structure of Na$_{2.7}$Ru$_4$O$_9$ by considering the observed superstructures, we begin from the intermediate-phase where a single super-lattice peak (\textbf{q}$_{1}$) appears (bottom panel of Fig.~\ref{fig:BVS}) and then two super-lattice peaks (\textbf{q}$_{1}$ and \textbf{q}$_{2}$) emerge in the room (low) temperature phase, which triples the cell-parameters $b$ and $c$ of the unit-cell (top panel of Fig.~\ref{fig:BVS}). The different types of RuO$_6$ octahedra of the corner sharing chains are comprised of single, double or triple edge-shared chains of RuO$_6$ octahedra. It then has irregular zigzag chains of RuO$_6$ octahedra along the crystallographic $c$-axis as well as large channel or cavities in the crystallographic $ac$-plane, in which multiple Na$^{+}$ atoms can reside (Fig.~\ref{fig:BVS}). Interestingly, along the crystallographic $b$-axis Na$_{2.7}$Ru$_4$O$_9$ forms a tunnel type structure, which is different from other prototype tunnel structures such as Rutile, Ramsdellite or Hollandite-type~\cite{Vogt1989,Carter2005} (see Section:~Tunnel structure).

Furthermore, the associated bond valence sum (BVS) calculation was performed using the \texttt{Fullprof} software suite~\cite{fullprof}, which reveals Ru$^{3+}$ (green RuO$_6$ octahedron) and Ru$^{4+}$ (light brown RuO$_6$ octahedron) coexisting in different valence states at both 300 and 450 K (Fig.~\ref{fig:BVS}). The electronic properties mainly depend on the charge, and charge separation is a common feature of charge ordering phenomena. At first sight, the emergence of the super-lattice peaks can be interpreted as the evidence of CO, such as in case of tunnel based structure’s like Hollandite-type~\cite{Carter2005} or in Na based compound Na$_x$CoO$_2$ has Co$^{3+}$/Co$^{4+}$ CO state ~\cite{Foo2004}. This features, therefore, indicates that Na$_{2.7}$Ru$_4$O$_9$ has unconventional CO without the loss of metallicity at room temperature. Above the first order phase transition at $T_{\textrm{C2}}$, it also shows the CO state but below the first order phase transition at ~$T_{\textrm{C2}}$ CO pattern drastically changed as shown in lower panel of Fig.~\ref{fig:BVS}.

\begin{figure}
\includegraphics[width=1\linewidth]{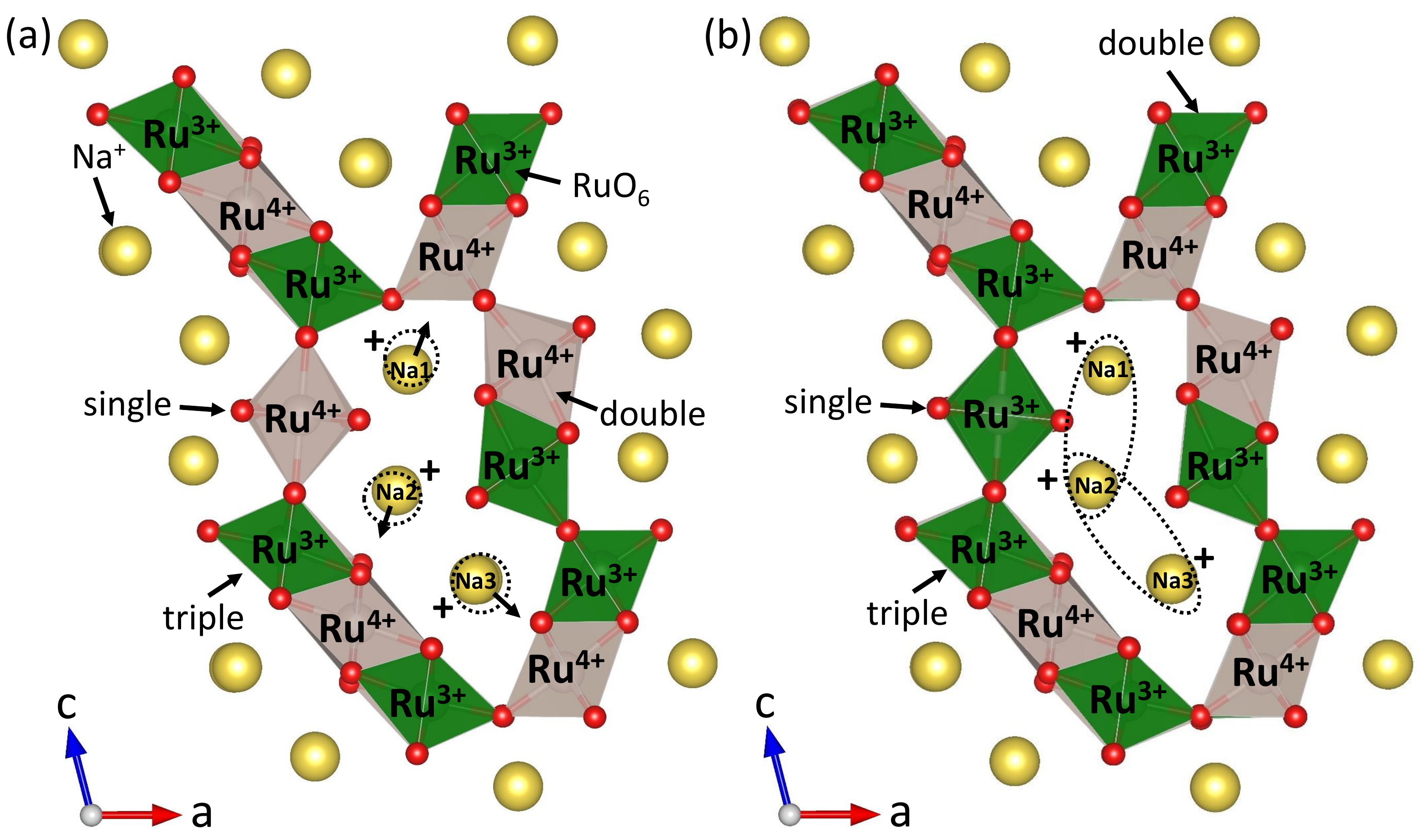}
\caption{\label{fig:CM}(Color online) Charge model based on the tunnel structure for Na$_{2.7}$Ru$_4$O$_9$ (a) below the first order transition ($\leq$~$T_{\textrm{C2}}$) at room temperature and (b) above the first order transition ($>$~$T_{\textrm{C2}}$) at 450 K.}
 \end{figure}

\begin{figure*}
\includegraphics[width=1\linewidth]{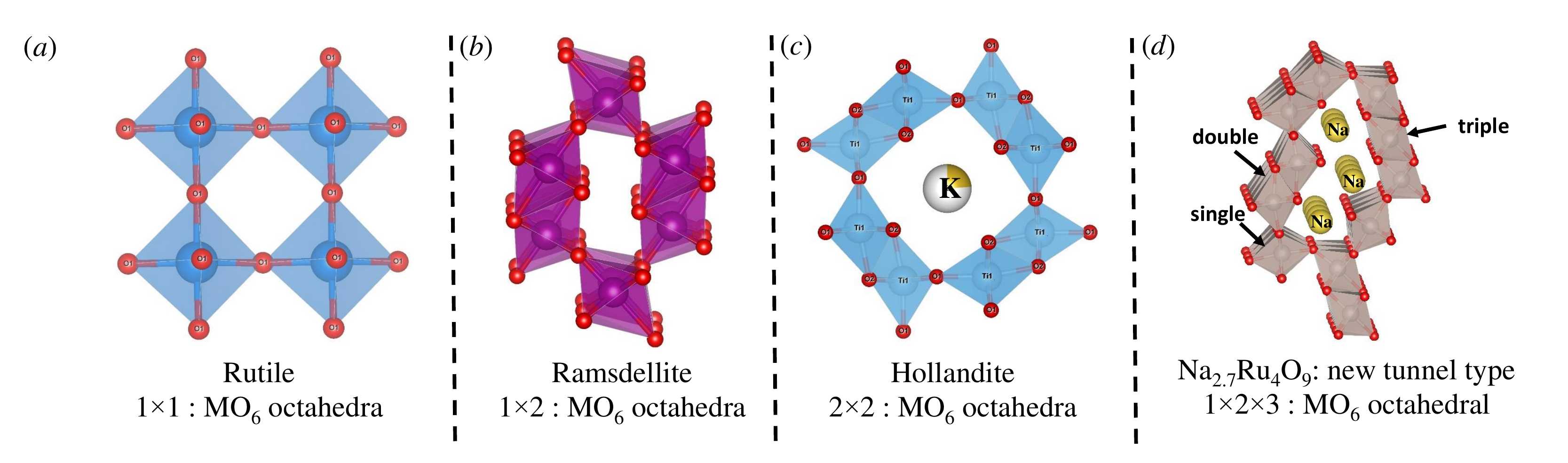}
\caption{\label{fig:tunnel-structures}(Color online) Comparison of prototype tunnel structures including Na$_{2.7}$Ru$_4$O$_9$ by MO$_6$ octahedra, where M is the transition metal in (a) Rutile, (b) Ramsdellite and (c) Hollandite type tunnels. (d) An enlarged view is shown of the new tunnel type structure for Na$_{2.7}$Ru$_4$O$_9$ formed by 1~$\times$~2~$\times$~3 edge shared single, double and triple chains.}
\end{figure*}

\begin{table*}
\caption{\label{tab:cell}
The unit-cell size and the space group of Na$_{2.7}$Ru$_4$O$_9$ for the HT-phase [695 K~$\equiv$~Prototype $C$2/$m$], the intermediate-phase [450 K~$\equiv$~$P$2/$m$] and the LT-phase [300 K~$\equiv$~Pseudo $P$2/$m$ ($P\overline{1}$)] have been investigated by single crystal x-ray diffraction measurements. The obtained values of the lattice parameters for different phases are listed below. The numbers in the parentheses are the respective error.
}
\begin{ruledtabular}
\begin{tabular}{cc@{\hspace{2em}}rrrr@{\hspace{2em}}c}
      &                   & &\multicolumn{2}{c}{lattice parameters} \\
\cline{2-6}
\\
Temperature  & $a$ ({\AA}) & $b$ ({\AA}) & $c$ ({\AA}) & $\beta$ ($^{\circ}$) & V ({\AA}$^{3}$)  \\
\hline
\\
695 K  &23.520 (11) &2.890 (1)   &10.953 (6)  &104.55(3)  &720.6 (6) \\
450 K  &23.311 (2)  &5.701 (4)   &11.057 (7)  &104.39(4) &1423.3(2)  \\
300 K  &23.342 (2)  &17.028 (16) &33.191 (3)  &104.43(7) &12776.1(6)   \\
\end{tabular}
\end{ruledtabular}
\end{table*}

\section{Discussion and Summary}
\subsection{Tunnel structure}
\label{sec:tunnel-structure}

The  crystal structure of Na$_{2.7}$Ru$_4$O$_9$ is composed of irregular zigzag chains forming tunnels along the $b$-axis. The Na ions are located inside these tunnels at three different crystallographic sites. However, a large number of Na$^{+}$ cations inside the tunnel experience strong mutual electrostatic repulsion and that increases as a function of temperature as explained through the charge model in Fig.~\ref{fig:CM} (a) and (b). Na$^+$ ions gain high mobility at high temperature that leads to Na$^{+}$ being disordered due to their weak bonding with ions forming a rigid framework~\cite{Kim2017}. The shift in the Na$^{+}$ NMR-spectra and the disappearance of one Na-site were observed for Na$_{3-x}$Ru$_4$O$_9$~\cite{Onodaa2000}. Therefore, in Na deficient Na$_{2.7}$Ru$_4$O$_9$, at or above the $T_{\textrm{C2}}$, one of the Na$^+$ atom moves and shares one of the Na atoms site in the tunnel as shown in Fig.~\ref{fig:CM} (b). The large displacements observed in the SC-XRD data along the $b$-axis favors ionic motions within the tunnels.

As Na$_{2.7}$Ru$_4$O$_9$ has tunnel geometry along the crystallographic $b$-axis with a large cross-section area, a larger number of alkali metal atoms can be accommodated than in Rutile, Ramsdellite and Hollandite as shown in Fig.~\ref{fig:tunnel-structures}. This feature makes it attractive in terms of sodium batteries~\cite{Yabuuchi2012} and ion motion related applications~\cite{Chen2015,Nikitina2017,Lekitsch2017,Kielpinski2002}. Interestingly, formation of the tunnel structure in Na$_{2.7}$Ru$_4$O$_9$ is purely of edge-sharing MO$_6$ octahedra chains (single (1$\times$1), double (1$\times$2), and triple (1$\times$3)), where M is Ru$^{3+}$/Ru$^{4+}$ and three cations Na$^{+}$ can reside in those tunnels/channels as shown in Fig.~\ref{fig:tunnel-structures} (d).

\subsection{Phase-transition}
\label{sec:phase-transition}

The possible prototype HT-phase ($a_0$ $\times$ $b_0$ $\times$ $c_0$) would be $C 2/m$ which is similar to the one reported by Maeno~\textit{et al.}~\cite{Maeno1994} as shown in Fig.~\ref{fig:structural-model} (c). A transition from this HT prototype phase $C 2/m$ ($\geq$ 630 K) to an intermediate-phase occurs (630 K $\leq T \leq$ 365 K). This intermediate phase transition leads to a doubling ($a_0$ $\times$ 2$b_0$ $\times$ $c_0$) of the unit-cell from the prototype structure (see Table~\ref{tab:cell}, Fig.~\ref{fig:structural-model} (b) and Fig.~\ref{fig:BVS}). Below the intermediate-phase, a first-order phase transition occurs as confirmed from the resistivity (Fig.~\ref{fig:FOT} (a)). Below 365 K new super-lattice peaks appear in the SC-XRD data, which makes the unit-cell 18 times ($a_0$ $\times$ 6$b_0$ $\times$ 3$c_0$) larger than the prototype phase unit-cell (Table~\ref{tab:cell}, Fig.~\ref{fig:structural-model} (b) and Fig.~\ref{fig:BVS})~\cite{Regan2005}.

These symmetry lowering structural-phase transitions enlarge the unit-cell drastically. Further, from the resistivity behaviour it is clear that Fermi surface of the Na$_{2.7}$Ru$_4$O$_9$ is comprised two types of electrons, 1) itinerant electrons (high spin state of Ru$^{3+}$/Ru$^{4+}$) responsible for the metallic behaviour and 2) localized electrons (possibly ultra-low spin state t$_{2g\uparrow}^2$ t$_{2g\downarrow}^2$ of Ru$^{4+}$) mainly responsible for the origin of the first order phase transition. Therefore, in such a case where electron states are localized, as in the Mott states, the observed \textbf{q}$_{1}$ and \textbf{q}$_{2}$ modulation in Na$_{2.7}$Ru$_4$O$_9$ may show CO-type modulations due to inter-site Coulomb interactions. However, localized charge-ordered states interact with itinerant electrons of Na$_{2.7}$Ru$_4$O$_9$ through SOC and retain the metallicity. IrTe$_2$ is one of the rare examples  showing similar behaviour with high SOC strength and unconventional CO. In this case, localized spin-orbit Mott states are assisted by Ir$^{4+}$ dimerizations~\cite{Ko2015}. Thus, the electronic behaviour of Na$_{2.7}$Ru$_4$O$_9$ is due to a subtle balance between itinerant and localized electrons~\cite{Subramanian2002}.

\subsection{Displacement pattern}
\label{sec:displaced-pattern}

\begin{figure*}
\includegraphics[width=0.91\linewidth]{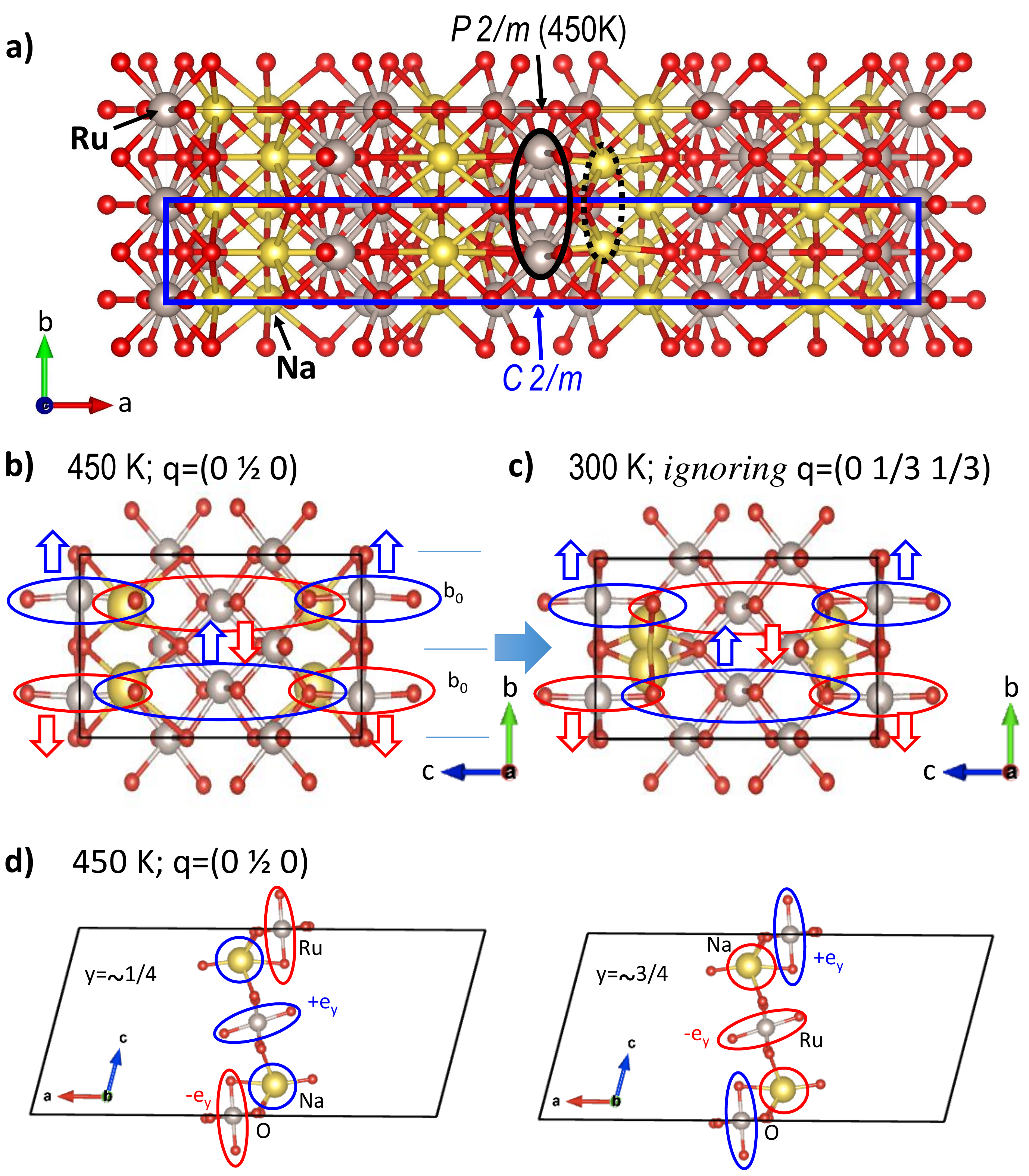}
\caption{\label{fig:Na-dis} (Color online) (a) The $P 2/m$ unit-cell of Na$_{2.7}$Ru$_4$O$_9$ at 450 K (\textbf{q}$_1$=(0, $\frac{1}{2}$, 0)). The reference solid blue line represents the HT prototype phase $C 2/m$  unit-cell. In the $P 2/m$ unit-cell, black solid and dashed rounded lines indicate the dimerization of Ru and Na ions along the crystallographic $b$-axis. The displacement pattern of Na ions at (b) 450 K and (c) 300 K in \textbf{q}$_1$=(0, $\frac{1}{2}$, 0) mode on the $bc$-plane projection, while (d) $ac$-plane at 450 K is shown at each $y$-layer, $1/4$ and $3/4$. The blue and red colours in (b-d) indicate positive and negative direction shifts.}
\end{figure*}

\begin{figure}
\includegraphics[width=1.0\linewidth]{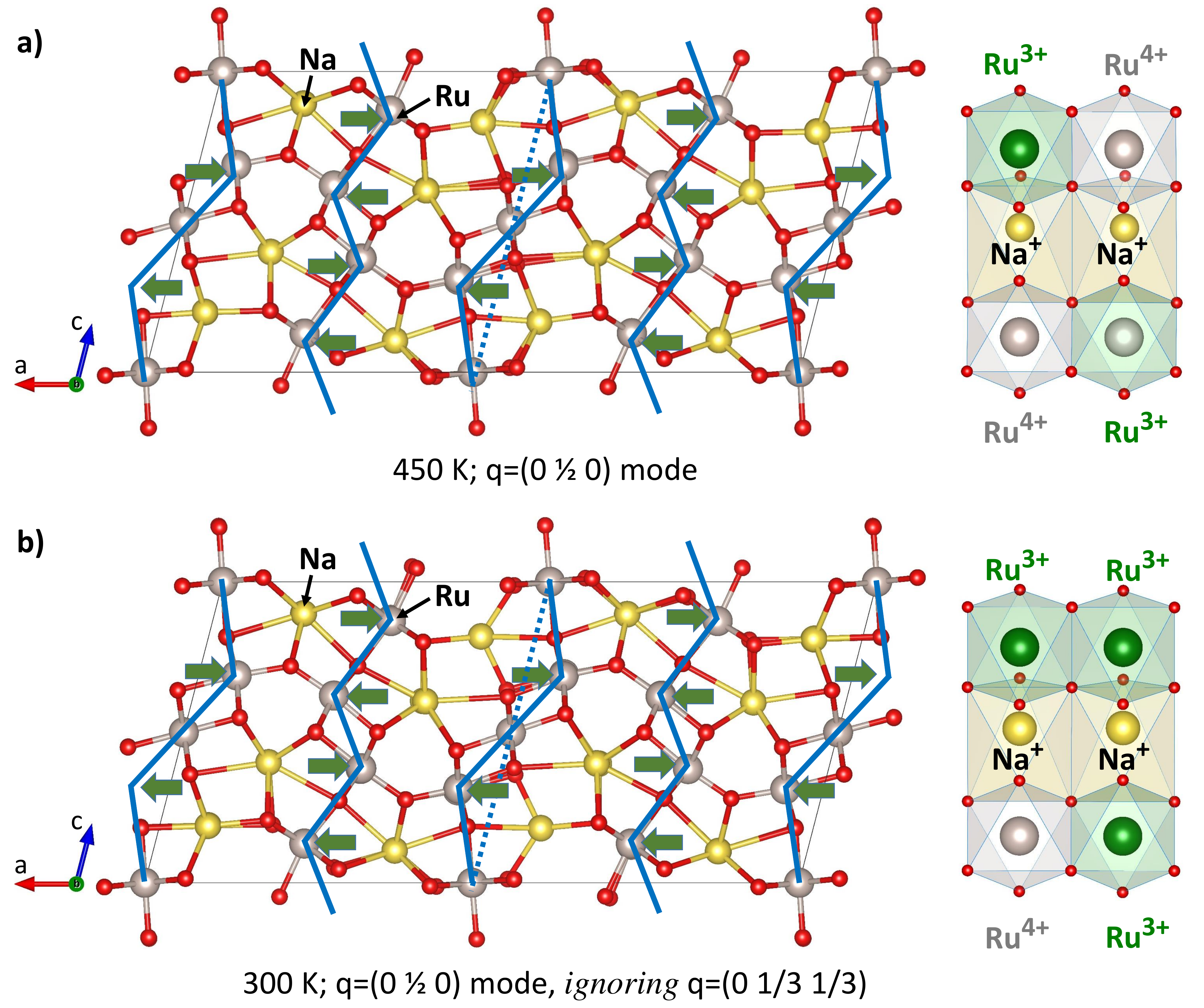}
\caption{\label{fig:Ru-dis} (Color online) The displacement patterns of Ru ions at (a) 450 and (b) 300 K with \textbf{q}$_1$=(0, $\frac{1}{2}$, 0) mode on the $ac$-plane projection. The right panel of the figures shows the highly distorted octahedrons local environment of displaced Na ions face sharing with different Ru$^{3+}$O$_6$ / Ru$^{4+}$O$_6$ octahedra at (a) 450 and (b) 300 K, respectively.}
\end{figure}

We now describe the structure and its displaced Ru and Na ions in more detail. Na ions are present at three different crystallographic sites inside the tunnels formed by RuO$_6$ octahedra, as shown in Fig.~\ref{fig:CM} (a) and (b). We observe a first-order phase transition with a strong anomaly in the electrical resistance and heat-capacity data. The superstructures have also been observed by SC-XRD for Na$_{2.7}$Ru$_4$O$_9$. This can be understood in terms of the Ru charge ordering, which has been confirmed by both temperature-dependent bulk and SC-XRD measurements (see Fig.~\ref{fig:FOT} and Fig.~\ref{fig:line-cut}). In order to avoid complexity we ignored the \textbf{q}$_2$ = (0, $\frac{1}{3}$, $\frac{1}{3}$) super-lattice reflections and analyzed the displacement-patterns by only considering the \textbf{q}$_1$ = (0, $\frac{1}{2}$, 0) superstructure mode with the same unit-cell size and the same space group $P$2/$m$. The displacement-patterns of Na and Ru are shown in Fig.~\ref{fig:Na-dis} (a-d) and Fig.~\ref{fig:Ru-dis} (a) and (b), respectively. It is worth to note that Ru and Na ions are dimerized in comparison with the HT prototype phase $C 2/m$ along the crystallographic $b$-direction as shown by the black solid and dashed rounded circles in Fig.~\ref{fig:Na-dis} (a).

From the intermediate-phase to the LT-phase, the structure shows large Na ion displacements, and one among the three Na sites is displaced along the $b$-direction by about $\frac{1}{4}$ from its normal site. The displaced RuO$_2$ and Na ions are shown by open arrows in Fig.~\ref{fig:Na-dis} (b) and (c) in the $a$-axis projection, which substantially changes the local environment of the structure. The observed large Na$^{+}$ ion displacements in Na$_{2.7}$Ru$_4$O$_9$ are in good agreement with Fig.~\ref{fig:Na-dis} (d), which shows the $b$-axis projection of the structure shown in Fig.~\ref{fig:Na-dis} (b) at 450 K. These Na ion displacements greatly affect the electronic structure of Na$_{2.7}$Ru$_4$O$_9$. This leads to displaced Ru ions along the $a$-direction, which forms zigzag chains along the crystallographic $c$-axis for both the intermediate and LT-phases as shown by the filled green arrow in Fig.~\ref{fig:Ru-dis} (a) and (b), respectively. We observed that due to the shift of the Na ions, Ru$^{4+}$ electrons are favoured in the intermediate-phase as shown in the right panel of Fig.~\ref{fig:Ru-dis} (a) and (b). However, Na octahedra sites are highly distorted face-sharing with both Ru$^{3+}$O$_6$ / Ru$^{4+}$O$_6$ octahedra and Ru-atoms form interactions with the alkali ions through oxygen, which stabilize $spd$ hybridization. This hybridization strongly influences the 4$d$ Ru electrons and the electronic properties of Na$_{2.7}$Ru$_4$O$_9$ are greatly affected accordingly. Although such face-sharing sites occur in both the intermediate and LT-phase structures, it is rather unusual to see such a site occupied by an alkali ion, Na, as we find in Na$_{2.7}$Ru$_4$O$_9$.

Therefore, the results of a structural investigation by the temperature-dependent SC-XRD provides the direct evidence for the formation of an unconventional charge ordering (CO). The system remains metallic, but parts of the Fermi surface may lose near the first order phase transition. This then explains the abrupt increase in the experimental resistivity at $T_{\textrm{C2}}$. Within this scenario the suppression of super-lattice peaks above $T_{\textrm{C2}}$ (Fig.~\ref{fig:line-cut} (c)) can be interpreted as Na$^{+}$ ion motions in the Na$_{2.7}$Ru$_4$O$_9$ lattice~\cite{Onodaa2000} being responsible for the increased localization of 4$d$ Ru$^{4+}$ electrons without the loss of metallicity. This scenario is very much consistent with magnetic susceptibility ($\chi$), where short range magnetic correlations were observed. We think that these results are responsible for the origin of the first-order phase transition and provide an unprecedented ionic displacement driven charge ordering in Na$_{2.7}$Ru$_4$O$_9$.

To conclude, we have presented evidence for a very unusual charge ordering in Na$_{2.7}$Ru$_4$O$_9$ at room temperature while retaining metallicity by combining the results of SC-XRD, electrical resistivity, specific heat, and susceptibility $\chi$ (\textit{T}). The temperature-dependent SC-XRD results show super-lattice peaks at \textbf{q}$_1$=(0, $\frac{1}{2}$, 0) and \textbf{q}$_2$=(0, $\frac{1}{3}$, $\frac{1}{3}$), clear evidence of symmetry breaking with large ionic displacements. This helps most probably establish charge ordering (CO) in Na$_{2.7}$Ru$_4$O$_9$. Na$_{2.7}$Ru$_4$O$_9$ hosts modest SOC of the Ru mixed valance heavy $d^4$ ions, which show several unique features such as CO in the metallic state and large alkali metal Na ion displacements in the tunnel lattice. More than one type of electron scattering is involved in the resistivity, alternatively making this material uniquely suitable example for several possible advanced applications such as in ion transport quantum computers and future energy storage materials. Further, the higher value of $\gamma$ indicates that Na$_{2.7}$Ru$_4$O$_9$ belongs to the class of strongly correlated electron systems.

\acknowledgments
We thank Daniel I. Khomskii and Sang-Wook Cheong for fruitful discussions. Work at the IBS CCES was supported by Institute for Basic Science (IBS) in Korea (Grant No. IBS-R009-G1)

\end{document}